\RequirePackage{etoolbox}
\csdef{input@path}{%
 {sty/}
 {img/}
}%
\csgdef{bibdir}{bib/}

\documentclass[ba]{imsart}
\pubyear{2020}
\volume{15}
\issue{3}
\doi{10.1214/19-BA1195}
\firstpage{1020}
\lastpage{1023}

\usepackage{amsthm}
\usepackage{amsmath}
\usepackage{natbib}
\usepackage[colorlinks,citecolor=blue,urlcolor=blue,filecolor=blue,backref=page]{hyperref}
\usepackage{graphicx}

\startlocaldefs
\numberwithin{equation}{section}
\theoremstyle{plain}

\endlocaldefs
\makeatletter
\g@addto@macro\normalsize{%
  \setlength\abovedisplayskip{5pt}
  \setlength\belowdisplayskip{5pt}
  \setlength\abovedisplayshortskip{1pt}
  \setlength\belowdisplayshortskip{1pt}
} \makeatother

\begin{document}

\begin{frontmatter}
\title{Discussion on ``Bayesian Regression Tree Models for Causal Inference: Regularization, Confounding, and Heterogeneous Effects'' by Hahn, Murray and Carvalho\thanksref{T1}}
\runtitle{Comment on Article by Hahn et al.}
\thankstext{T1}{Main article \href{https://projecteuclid.org/euclid.ba/1580461461}{DOI: 10.1214/19-BA1195.} }

\begin{aug}
\author{\fnms{Liangyuan} \snm{Hu}\thanksref{addr1}\ead[label=e1]{liangyuan.hu@mssm.edu}}

\runauthor{Hu L}

\address[addr1]{Department of Population Health Science and Policy, Icahn School of Medicine at Mount Sinai,
    \printead{e1} 
}

\end{aug}

\end{frontmatter}

Congratulations to Hahn, Murray and Carvalho on a nice contribution. The authors propose the Bayesian causal forest (BCF) model. It reduces the bias  and improves frequentist coverage of Bayesian credible intervals of treatment effect estimates in the presence of strong confounding and treatment effect heterogeneity. The BCF model builds upon the popular and empirically proven prediction method, Bayesian Additive Regression Trees (BART) \citep{chipman2010bart}. BCF reformulates the response surface model as the sum of two functions, one modeling the prognostic impact of the control variables and one representing the treatment effect. This formulation directly incorporates estimates of the propensity score (PS) which induces a covariate-dependent prior on the regression function and regularizes treatment effect heterogeneity separately from the prognostic effect of control variables. It is impressive that the proposed method, in many settings, has better bias reduction, more consistent 95\% coverage probability and shorter uncertainty intervals compared to the vanilla BART, which boasts better performance in a host of modern causal inference studies, including  \cite{hill2011bayesian}, \cite{wendling2018comparing}, \cite{dorie2019automated} and \cite{hu2020estimation}, to name a few. 

This paper offers an extensive study to explicate and evaluate the performance of the BCF model in different settings and provides a detailed discussion about its utility in causal inference. It is a welcomed addition to the causal machine learning literature. 

I will emphasize the contribution of the BCF model to the field of causal inference through discussions on two topics: 1) the difference between the PS in the BCF model and the Bayesian PS in a Bayesian updating approach, 2) an alternative exposition of the role of the PS in outcome modeling based methods for the estimation of causal effects. I will conclude with comments on avenues for future research involving BCF that will be important and much needed in the era of Big data. 

\section{Distinction from the Bayesian propensity scores}
It is necessary to make a distinction between incorporating the estimated PSs as a covariate in the BCF model and combining so called ``Bayesian propensity scores'' and Bayesian inference for the estimation of causal effects in a single Bayesian updating approach \citep{zigler2013model,zigler2014uncertainty}. The Bayesian PS has received recent attention in the literature \citep{kaplan2012two,zigler2016central,liao2020uncertainty}. In essence, a series of work have demonstrated that the model feedback, i.e.,  the propagation of information from the outcome model to the PS model, would distort inferences about the causal effect. In the BCF model, the independent BART prior is placed over $f$, $f \sim \text{BART} (X,Z, \hat{\pi})$, where $\hat{\pi}$ is the estimated PS and included as one of the splitting dimensions.  The $\hat{\pi}$ is included in the BCF model as an additional covariate for splitting and is not updated or contaminated by the outcome information, thereby setting BCF free from the inference issue caused by the Bayesian model feedback.

\section{The role of the propensity score: connections to the confounding function}
I now turn to exposing the role of the PS in reducing the bias in the estimates of causal effects under targeted selection. The inclusion of the PS as a covariate in the response model is closely connected to the confounding function approach \citep{robins1999association}, designed for removing the bias due to unmeasured confounding from the treatment effect estimates.  The targeted selection described in \cite{hahn2020bayesian} suggests that the treatment assignment probability $\pi(x)$ depends on $\mu(x) = E(Y|Z=0,x)$. Using Figure 4 as an example to illustrate, individuals who would have higher outcomes without treatment are more likely to be treated. In another word, treated individuals would have higher (potential) outcomes than untreated individuals to no treatment. This is a violation of the ignorability assumption. To see why,  define a confounding function as $c(z,x) = E(Y(z)|Z=1, X=x) - E(Y(z) | Z=0, X= x)$, $z \in \{0,1\}$. When ignorability holds, $c(z,x)=0$ for $z=0$ and $z=1$. When ignorability is violated, such as in the presence of targeted selection,  $c(0,x)>0$. The violation of ignorability gives rise to biased estimates of the causal effects. 
An unbiased effect estimate can be obtained by ``correcting'' the \emph{observed} outcome for unmeasured confounding, namely, $Y- [E(Y(z) |Z=z, x) - E(Y(z) |x)]$ \citep{robins1999association, brumback2004sensitivity}.  Applying the law of total expectation to $E(Y(z) |x)$ yields a ``corrected'' outcome,
\begin{eqnarray}\label{eq:correction}
Y^C= Y-\lbrace(1-\pi(x))z - \pi(x)(1-z)\rbrace c(z,x), \;  z\in \{0,1\}.
\end{eqnarray} 
Differencing the conditional expectations of $Y^C$ between the treated and untreated individuals gives an unbiased effect estimate, $E(Y^C|X=x, Z=1) - E(Y^C|X=x,Z=0)$. In equation~\eqref{eq:correction}, $c(z,x)$ is a user-supplied prior distribution (a range of values in frequentist approaches) representing our beliefs about the degree of ignorability violation, or, targeted selection \citep{hogan2014}. We see that the PS, $\pi(x)$, is an integral part in the outcome for causal modeling when there is a need to remove the bias attributable to targeted selection. Relating to the inclusion of $\hat{\pi}$ in the BCF model, the $c(z,x)$ can be deemed as characterizing how big of a role the PS plays in the estimation of causal effects. With strong targeted selection, $c(z,x)$ would deviate from zero substantially, then $\hat{\pi}$ is important for bias reduction. In the absence of targeted selection, $c(z,x)$ would be close to zero, and the role of $\hat{\pi}$ is diminished.


\section{Final thought on possible extensions }
My final thought is about extending the BCF model to meet the emerging methodological needs, particularly in the Biostatistical research. First, a common causal estimand of interest is the average treatment effect on the treated (ATT).  The BCF model does not seem to be readily implementable for the ATT estimation. Although, the idea of ps-BART can be easily applied to estimate the ATT effect. Second, in the era of Big data,  given the wealth of information captured in large-scale data, it is rare that treatment regimens are defined in terms of two treatments only. Refined causal inference approaches are in great demand for the multiple treatment settings. \cite{hu2020estimation} investigated the operating characteristics of several machine learning based causal inference techniques in the multiple treatment setting, and found that BART-based methods generally had the best performance. It would be a useful addition to the causal machine learning literature if the BCF model can be extended to simultaneously compare more than two active treatments.

%
%
%

\bibliographystyle{ba}
\bibliography{comment}

\end{document}